# High-dynamic-range imaging of nanoscale magnetic fields using optimal control of a single qubit

T. Häberle[1]*, D. Schmid-Lorch[1], K. Karrai[2], F. Reinhard[1] and J. Wrachtrup[1]

[1] 3. Physikalisches Institut and Stuttgart Research Center of Photonic Engineering (SCoPE), Universität Stuttgart, Pfaffenwaldring 57, Stuttgart D-70569, Germany

[2] attocube systems AG, Koeniginstrasse 11a, Munich D-80539, Germany

We present a novel spectroscopy protocol based on optimal control of a single quantum system. It enables measurements with quantum-limited sensitivity ($\eta_\omega \propto \frac{1}{\sqrt{T_2^*}}$, $T_2^*$ denoting the system's coherence time) but has an orders of magnitude larger dynamic range than pulsed spectroscopy methods previously employed for this task. We employ this protocol to image nanoscale magnetic fields with a single scanning NV center in diamond. Here, our scheme enables quantitative imaging of a strongly inhomogeneous field in a single scan without closed-loop control, which has previously been necessary to achieve this goal.

Optimal control of quantum systems is an experimental technique that has evolved over the two past decades [1–3] as a generalization of related techniques like composite pulses [4] or adiabatic control [5]. It implements unitary operations ("quantum gates") of very high fidelity by irradiating a quantum system with numerically optimized excitation pulses. Amplitude and phase of this pulse are an arbitrary function of time, which is tailored such as to result in a specific unitary operation.

Numerical optimization can generate pulses that achieve near-perfect operation (i.e. high fidelity) over a wide range of experimental parameters, such as excitation power or detuning, rather than a single specific set. This is in contrast to simple (e.g. rectangular) pulses and arises from the fact that optimization has access to the much larger space of arbitrary amplitude and phase profiles. Thanks to these additional degrees of freedom, the resulting pulse can satisfy a larger number of constraints. In practice, this has been used to generate "robust" pulses which are immune against fluctuations of the excitation power, or pulses that implement a specific operation within a large bandwidth of different system frequencies [3], as they may arise e.g. by inhomogeneous broadening.

Here we show that optimal control can be used to achieve an opposite goal, a pulse that is *maximally sensitive* to fluctuations of one experimental parameter (in our case the static magnetic field) while it preserves robustness against fluctuations of all other parameters and, in particular, a large operating bandwidth. With these properties, such a pulse enables sensitive spectroscopy of a system even in the presence of large unknown frequency offsets.

The concept is illustrated in more detail in Fig. 1(a-b). Sensitive spectroscopy classically relies on sharp selective excitation (Fig. 1(a)), realized for instance by a long low-power excitation pulse (Rabi spectroscopy) or a suitable pulse sequence (Ramsey spectroscopy). We extend these schemes by designing an optimal control pulse which generates a grating of equally sharp excitation lines, evenly spaced over a large bandwidth (Fig. 1(b)). With this protocol, small changes of the system's resonance frequency can be tracked without tuning the excitation pulse to the system's frequency.

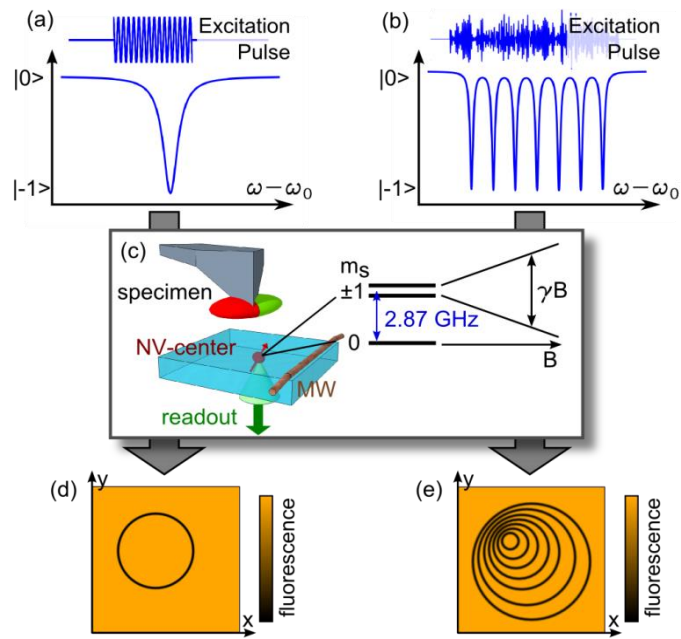

FIG 1

Measurement scheme: (a-b) Precision spectroscopy of a two-level system (level spacing $\omega_0$). Conventional Rabi spectroscopy realizes a highly frequeny-selective excitation by a weak, rectangular excitation pulse (a). Optimal control pulses (b, top half) can be tailored to generate multiple such resonance dips (lower half), thereby increasing bandwidth while retaining sensitivity. (c) Measurement geometry: a magnetic specimen is scanned by an AFM over an NV-center in bulk diamond, inducing a spatially varying Zeeman splitting (right half). A wire allows for microwave (MW)-spectroscopy of the NV-center, whose spin state can be read out optically from below. (d) and (e): Expected fluorescence images for excitation with pulses (a) and (b), respectively. Every resonance dip translates into a dark resonance ring in the fluorescence image, corresponding to a contour line of the specimen's magnetic field.

We apply this scheme to the imaging of nanoscale magnetic fields with a single scanning nitrogen-vacancy (NV) center in diamond [6–9]. This recently developed technique promises to enable magnetic field imaging with a sensitivity comparable or better than other methods (e.g. Hall sensors or MFM), but with a vastly superior spatial resolution [10,11]. It images the magnetic field of a magnetic nanostructure by mapping the position-dependent Zeeman shift of a single defect center. This center is scanned over the structure (or, as in our case, vice-versa) by an atomic-force microscope (Fig. 1(c)). The Zeeman shift is typically measured by selectively exciting the transition $|0\rangle \rightarrow |-1\rangle$ of the triplet spin ground state with a microwave pulse and reading out population of the $|0\rangle$ spin state optically. In this approach, a single contour line of the magnetic field is revealed as a dark "resonance fringe" where the magnetic field tunes the system into resonance with the microwave and fluorescence drops (Fig. 1(d)). The sensitivity of this scheme is quantum-limited (i.e.: limited by the intrinsic linewidth of the system $1/T_2^*$). However, being restricted to a single contour line, it is unable to quantitatively map the magnetic field in a single run. This problem has been solved by lock-in techniques, which continuously tune the microwave into resonance as the center scans through a varying magnetic field [9]. Unfortunately, these techniques are technically delicate to handle and irreversibly lose lock when the center scans over large field gradients or blinks.

We combine the advantages of both approaches by exciting the center with the protocol of Fig. 1(b). This excitation grating translates into multiple resonance fringes and hence allows recording of several contour lines of the magnetic field in a single scan (Fig. 1(e)). The magnetic field can be quantitatively reconstructed by post-processing of this data, without any feedback and closed-loop control in the actual experiment. Sections of invalid data (caused, e.g. by blinking of the center) can be removed without damaging valid regions of the scan and the sensitivity can still be pushed to the quantum limit by choosing a sufficiently fine excitation grating. This approach is the core idea of this paper, and all the following will be a more detailed report on its experimental realization.

We numerically optimized a "grating pulse" as described above using the GRAPE algorithm [3,12]. This algorithm accepts as an input an initial non-optimal pulse and an arbitrary, frequency-dependent pattern of target magnetization [13]. It returns an arbitrary-waveform pulse that is optimized to excite this pattern. We exploit this freedom by supplying the algorithm with a target-grating of several (to increase bandwidth), sharp (to retain high sensitivity), regularly spaced excitation lines (Fig. 2(a)). Hence, the resulting excitation pattern is maximally sensitive to small variations of the system's frequency within a wide bandwidth, while optimal control guarantees robustness against fluctuations in other parameters such as MW power.

We started GRAPE optimization from a pulse that had been manually optimized to fit the target pattern coarsely. We found this initial guess to influence the number of iterations needed as well as the overall power-dissipation of the resulting pulse, but not its fidelity.

The resulting pulse is illustrated in Fig. 2(b-c). We experimentally verified its performance by spectroscopy on a single NV center (Fig. 2(d)). Here, the pulse was frequency-shifted by modulating it with a carrier frequency that was swept over the resonance frequency of the NV center, simulating a varying magnetic field.

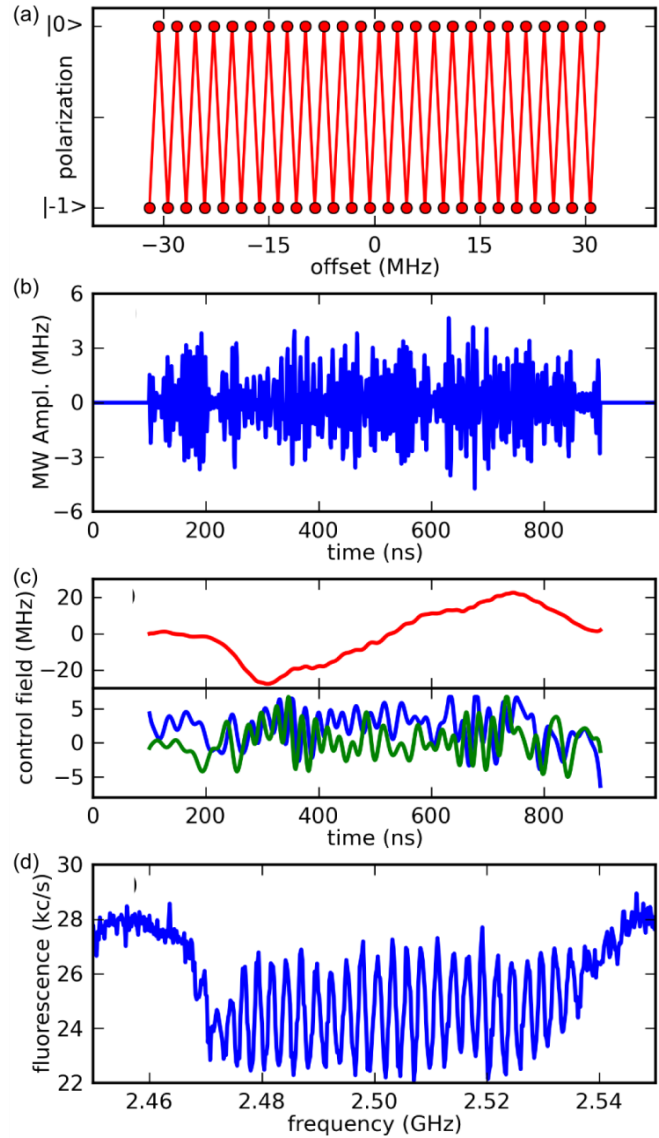

FIG 2

Pulse optimization: (a) Target magnetization for the optimization algorithm. (b,c) Optimized pulse as it was calculated by the GRAPE algorithm. (b) waveform of the pulse, (c) quadrature envelopes x (blue) and y (green) and detuning z (red) of the same pulse. (d) Experimental verification: spectroscopy of a single NV center using pulse (b,c). The grating pattern is clearly visible.

We used this spectroscopy pulse to acquire two-dimensional maps of magnetic fields in the geometry of Fig. 1(c). As a convenient test sample, we employed a commercial magnetic force microscopy (MFM) tip (Bruker MESP). By simultaneously scanning this sample and performing optimal-control spectroscopy on the NV-center, we obtain the NV fluorescence as a function of sample position shown in Fig. 3(a). Here, every dip of Fig. 2(d) translates into a dark fringe, corresponding to a contour line of the sample's magnetic field.

We reconstructed a two-dimensional map of the magnetic field by post-processing of this data (Fig. 3(b)). We computed an initial guess by manually assigning a magnetic field to each fringe and subsequently performing a least-square fit to best match the observed fluorescence (Fig. 3(a)). We note that this manual assignment is in principle ambiguous, since every fringe could be assigned to any dip of the excitation grating. However, this problem appears surmountable. By removing an individual dip from the excitation grating, a "missing fringe" could uniquely identify a particular magnetic field. Also, a second image could be acquired with an excitation grating shifted by a fraction of the grating's frequency spacing. In this way, the gradient of the magnetic field could be obtained along with the contour lines, hence allowing for a unique assignment of the fringes. Notably, the use of post-processing enables us to discard invalid portions of the image. In our case, this is visible as the "uncharted territory" on the reconstructed field distribution, corresponding to areas outside the bandwidth of the spectroscopy pulse.

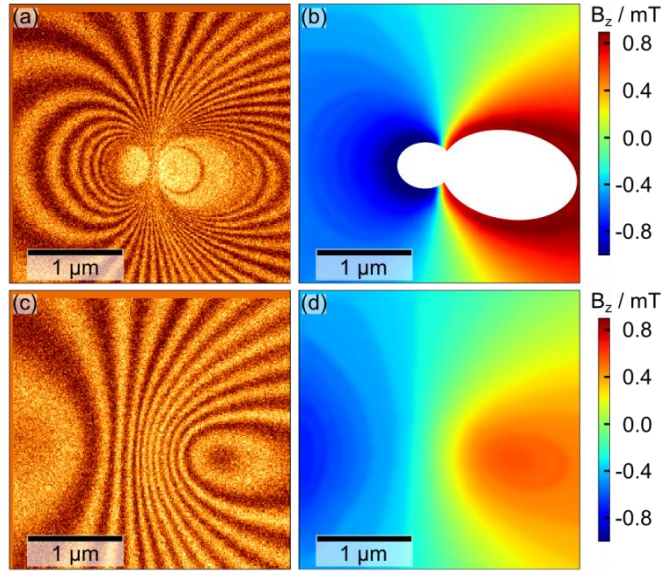

FIG 3

Resonance fringe (contour line) images and reconstructions of the magnetic field of the MFM-tip: (a) Resonance fringes taken in contact mode using excitation pulse Fig. 2(b,c). (b) Reconstruction of the underlying magnetic field distribution (projection $B_z$ along z-axis of the NV-center) from (a). (c) Resonance fringes when the tip is lifted by 600 nm. (d) Corresponding reconstruction image of (c). For both reconstructions a bias field of 7.8 mT was subtracted from the field values.

We now turn to a discussion of the performance of our measurement method. Fundamentally, a spectroscopic measurement is limited by the intrinsic linewidth of the system, in our case $\Gamma = 1/T_2^*$, $T_2^*$ denoting the NV center's coherence time. This limit is commonly referred to as the "standard quantum limit". We demonstrated that our method operates at this limit by analyzing the contrast of our images for varying spacing of the excitation grating (Fig. 4(a)). Indeed, we observe a reduced contrast for a decreasing spacing of the grating, where convolution of the excitation pattern (Fig. 2(a)) with the intrinsic linewidth of the system smears out the signal. We quantitatively compared this decay to the contrast of a free-induction-decay (FID) measurement (Fig. 4(c)), finding excellent agreement.

Having established the quantum-limited performance of our measurement method, we can now compute the optimal point of operation. The sensitivity $\eta$ of our measurement is a tradeoff between a small spacing of the grating (leading to a higher resolution) and $T_2^*$-decay (leading to a reduced contrast for decreasing spacing). Quantitatively, this is expressed as [11]

$$\eta = \frac{\Delta}{c_0 \cdot \exp\left(-\frac{1}{2\Delta^2 T_2^{*2}}\right)} \cdot \frac{1}{\sqrt{S_0 \cdot D}} \cdot \frac{1}{\gamma}.$$

Here, $\gamma = 28$ MHz/mT is the NV gyromagnetic ratio. The term $\sqrt{S_0 \cdot D}$ models photon shot noise, with $S_0$ denoting the photon count rate and $D$ the duty cycle ($D = T_{readout}/T_{seq}$, readout time as a fraction of total measurement time). The fringe contrast is expressed by the first term, where $c_0$ denotes the maximum achievable contrast, $\Delta$ the grating's spacing, and the exponential describes the $T_2^*$-limited decay of contrast found above. The resulting sensitivity (Fig. 4(c), green line, mind the inverted right-hand side y-axis) indeed has an optimum at $\frac{1}{\Delta} = T_2^*$, where our NV-center with $T_2^* = 416$ ns reaches a sensitivity of $\eta = 4.50$ µT/$\sqrt{\text{Hz}}$. ($c_0 = 0.3$, $S_0 = 150$ kHz, $T_{readout} = 300$ ns, $T_{seq} = 4100$ ns)

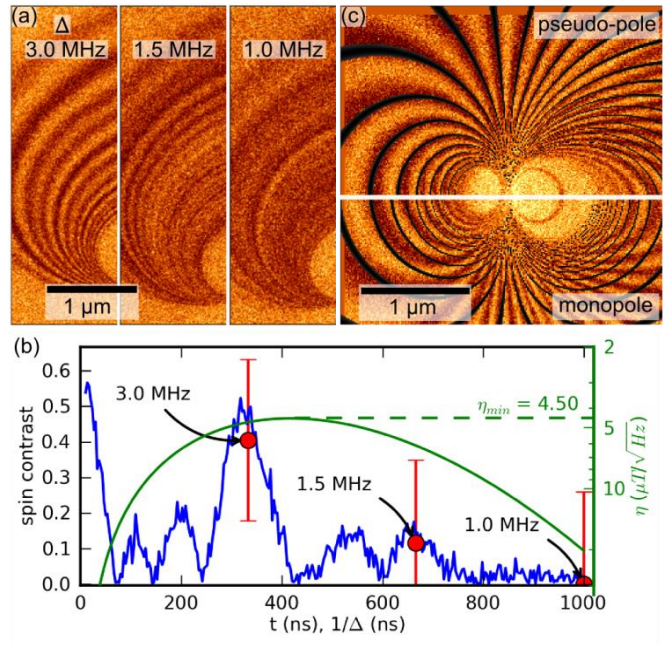

FIG 4

(a,b): Verification of quantum-limited sensitivity. (a) Spectroscopy of an NV-center for different spacings $\Delta$ of the employed excitation grating. For decreasing spacing, contrast decays due to the center's intrinsic linewidth $\Gamma = \frac{1}{T_2^*}$. (b) FID-measurement (blue line) and image contrast (red dots) from (a), as a function of an "effective pulse duration" $t_{\text{eff}} = \frac{1}{\Delta}$. Green line: estimate of the photon-shot-noise-limited sensitivity as a function of $t_{\text{eff}}$, mind the inverted right-hand side y-axis. The optimal sensitivity of $\eta_{min} = 4.50$ µT/$\sqrt{\text{Hz}}$ is marked by the dashed line. (c) Analysis of the magnetic field maps. The magnetic field of the MFM tip distinctly differs from a magnetic monopole (lower half) but is well described by a recently introduced pseudopole model ( [14] upper half).

We finally use the quantitative field maps obtained by this technique to benchmark several analytical field models for magnetic force microscopy tips. We find that our data is not well described by approximating the tip by a magnetic monopole. This approximation is widely used [15–20], motivated by the fact that the dipole density of the magnetic coating should closely resemble a strong magnetic charge on the tip apex, but is a coarse approximation [21]. Our data is better described by the more elaborate pseudo-pole model [14], where the field decays proportional to $1/R$, ($R$ denoting distance to the tip), in contrast to the $1/R^2$ dependence of a monopole. We estimate that the remaining discrepancies can be explained by shape anisotropy of the tip which is not considered in the model. This insight is of interest for MFM studies, since an analytical model of the tip's field is a crucial ingredient for quantitative measurements, but its experimental validation has been difficult so far.

In summary, we have used optimal control to create pulses that are maximally sensitive to fluctuations in the magnetic field over a wide range of frequencies. The use of optimal control allows manipulating magnetization over a bandwidth of $\Delta\omega = \tau\Omega_R^2$ ($\tau$ denoting the total length of the spectroscopy pulse, $\Omega_R$ the maximum admissible Rabi frequency). This range can be orders of magnitude larger than the range accessible by pulsed methods such as Ramsey spectroscopy, which are in general limited to $\Delta\omega = \Omega_R$. The resolution of our scheme is on par with pulsed methods and equally quantum-limited by the system's intrinsic linewidth $\Gamma = 1/T_2^*$. In our case, this leads to a sensitivity of $\eta = 4.50\ \mu\text{T}/\sqrt{\text{Hz}}$ and a dynamic range of more than 2.2 mT.

We have applied this method to the two-dimensional mapping of magnetic fields, where it enables the acquisition of multiple contour lines of the field in a single scan. Moreover, we were able to extract a quantitative two-dimensional map of the magnetic field by post-processing, a result which has hitherto required the use of experimentally challenging lock-in techniques [9]. Our method has proven to be robust, since sections of invalid data can be removed during post-processing, and, consequently, even large magnetic field gradients can be imaged. Capitalizing on this capability, we have validated an analytical model for the field of magnetic-force microscopy tips [14].

We believe that our method will equally find applications in the spectroscopy of stationary (non-scanning) quantum systems such as magnetometers or atomic clocks. Here, it promises to further improve the dynamic range of the recently demonstrated phase-estimation algorithms (PEA) [22,23], which are currently limited to a bandwidth on the order of the Rabi frequency $\Delta\omega = \Omega_R$.

We dedicate this work to the memory of Carsten Georgi, who greatly contributed to this project. We acknowledge financial support by the Max-Planck-society, the EU (Squtec), Darpa (Quasar), BMBF (CHIST-ERA) and contract research of the Baden-Württemberg foundation.

*t.haeberle@physik.uni-stuttgart.de